\documentclass[prl,aps,epsfig,superscriptaddress,showpacs,twocolumn,notitlepage]{revtex4-1}
\usepackage{amsmath,amssymb,amsfonts,amstext}
\usepackage{graphicx,breakurl,color,setspace}
\usepackage[T1]{fontenc}
\usepackage[latin9]{inputenc}
\usepackage[normalem]{ulem}
\usepackage[unicode=true,pdfusetitle,pdfborder={0 0 1}]{hyperref}
\usepackage{upgreek}
\usepackage{bm}
\usepackage{mathrsfs}

\hypersetup{colorlinks,
linkcolor=blue,          citecolor=blue,        filecolor=blue,      urlcolor=blue}

\usepackage{times}

\graphicspath{{../figures/}}

\begin{document}

\title{Experimental Test of Leggett's Inequalities with Solid-State Spins}

\author{Xianzhi Huang}\thanks{These authors contributed equally to this work.}
\author{Xiaolong Ouyang}\thanks{These authors contributed equally to this work.}
\author{Wenqian Lian}\thanks{These authors contributed equally to this work.}
\author{Wengang Zhang}\author{Xin Wang}\author{Huili Zhang}\author{Yefei Yu}\author{Li He}\author{Yanqing Liu}\author{Xiuying Chang}\author{Dong-Ling Deng}\email{dldeng@tsinghua.edu.cn}\author{Luming Duan}
\email{lmduan@tsinghua.edu.cn}

\affiliation{Center for Quantum Information, IIIS, Tsinghua University, Beijing 100084, PR China}

\begin{abstract}
Bell's theorem states that no local hidden variable model is compatible with quantum mechanics.  Surprisingly, even if we release the locality constraint, certain nonlocal hidden variable models, such as the one proposed by Leggett, may still be at variance with the predictions of quantum physics. Here, we report an experimental test of Leggett's nonlocal model with solid-state spins in a diamond nitrogen-vacancy center. We entangle an electron spin with a surrounding weakly coupled $^{13}C$ nuclear spin and observe that the entangled states violate Leggett-type inequalities by more than four and seven standard deviations for six and eight measurement settings, respectively. Our experimental results are in full agreement with quantum predictions and violate Leggett's nonlocal hidden variable inequality with a high level of confidence.
\end{abstract}

\maketitle
Realism and locality are two fundamental concepts in classical physics \cite{Einstein1935Can,Bell1964OnEPR,Brunner2014Bell}. Roughly speaking, locality requires that events happened in space-like separated regions can not influence each other,  while realism suggests that the results of observations are predetermined by the intrinsic properties of a physical system and should be independent of the measurement \cite{leggett2008realism}.  Quantum physics, however, challenges these concepts in a profound way---no hidden-variable theory based on the joint assumption of realism and locality can reproduce all quantum correlations \cite{Bell1964OnEPR,Brunner2014Bell}. This fact is now well-known through Bell's theorem \cite{bell2004speakable}  and has been verified by a number of experiments with different platforms \cite{Freedman1972Experimental,Aspect1982Experimental,Weihs1998Violation,Rowe2001Experimental,
Giustina2013Bell,Christensen2013Detection,Eibl2003Experimental,Zhao2003Experimental,
Lanyon2014Experimental,Hofmann2012Heralded,Pfaff2013Demonstration,Ansmann2009Violation,Zu2013Experimental,
Hensen2015Loophole,Giustina2015Significant,Shalm2015Strong,Rosenfeld2017Event}. In particular, recent experiments with entangled electron spins \cite{Hensen2015Loophole},  photons \cite{Giustina2015Significant,Shalm2015Strong}, and atoms \cite{Rosenfeld2017Event} have been reported to close the detection and locality loophole simultaneously. This, together with the Bell Test project that attempts to close the freedom-of-choice loophole \cite{big2018challenging}, has reasonably established the violation of local realism in quantum physics a validated fact.

Then, should non-local realism be consistent with quantum physics? This is a natural question, but the answer is complicated. On the one hand, Bohm's interpretation \cite{Bohm1952Asuggested} of quantum mechanics clearly implies that certain non-local hidden variable (NLHV) models can indeed reproduce all predictions of quantum physics. On the other hand, however, there also exist other NLHV models that are proved to be incompatible with quantum predictions. The first testable example of such NLHV model was the one proposed by Suarez and Scarani \cite{Suarez1997Does}, which has been falsified in a series of subsequent experiments \cite{Zbinden2001Experimental,Stefanov2002Quantum}.  Another notable example involves the one introduced by Leggett \cite{leggett2003nonlocal}.  This NLHV model fulfills the so-called Leggett's inequalities, but quantum correlations can violate them.  Leggett's model has attracted  considerable attentions in the community and a number of experiments have been carried out to test it \cite{groblacher2007experimental,paterek2007experimental,branciard2007experimental,
branciard2008testing,Paternostro2010Testing,romero2010violation,Lee2011Faithful,Cardano2013Violation,
hasegawa2012falsification}. All these experiments support the predictions of quantum mechanics and show violations of Leggett's inequalities. Nevertheless, most of these experiments use photons \cite{groblacher2007experimental,paterek2007experimental,branciard2007experimental,
branciard2008testing,Paternostro2010Testing,romero2010violation,Lee2011Faithful,
Cardano2013Violation} and Leggett's model has never been tested in a solid-state system hitherto, which is in sharp contrast to the case for Bell inequalities. Given the important roles solid-state systems play in quantum information sciences, it is highly desirable that the Leggett inequalities should also be tested in such systems. Violation of Leggett's inequality requires preparation of entangled states with a very high fidelity and correlation measurements in various complementary settings, which are experimentally challenging. Therefore, test of Leggett's inequality, apart from its fundamental interest, is also a demonstration of good quality of entanglement control in the corresponding quantum systems.

\begin{figure}[t]
	\includegraphics[trim=0cm 0cm 0cm 0cm, clip,scale=0.36]{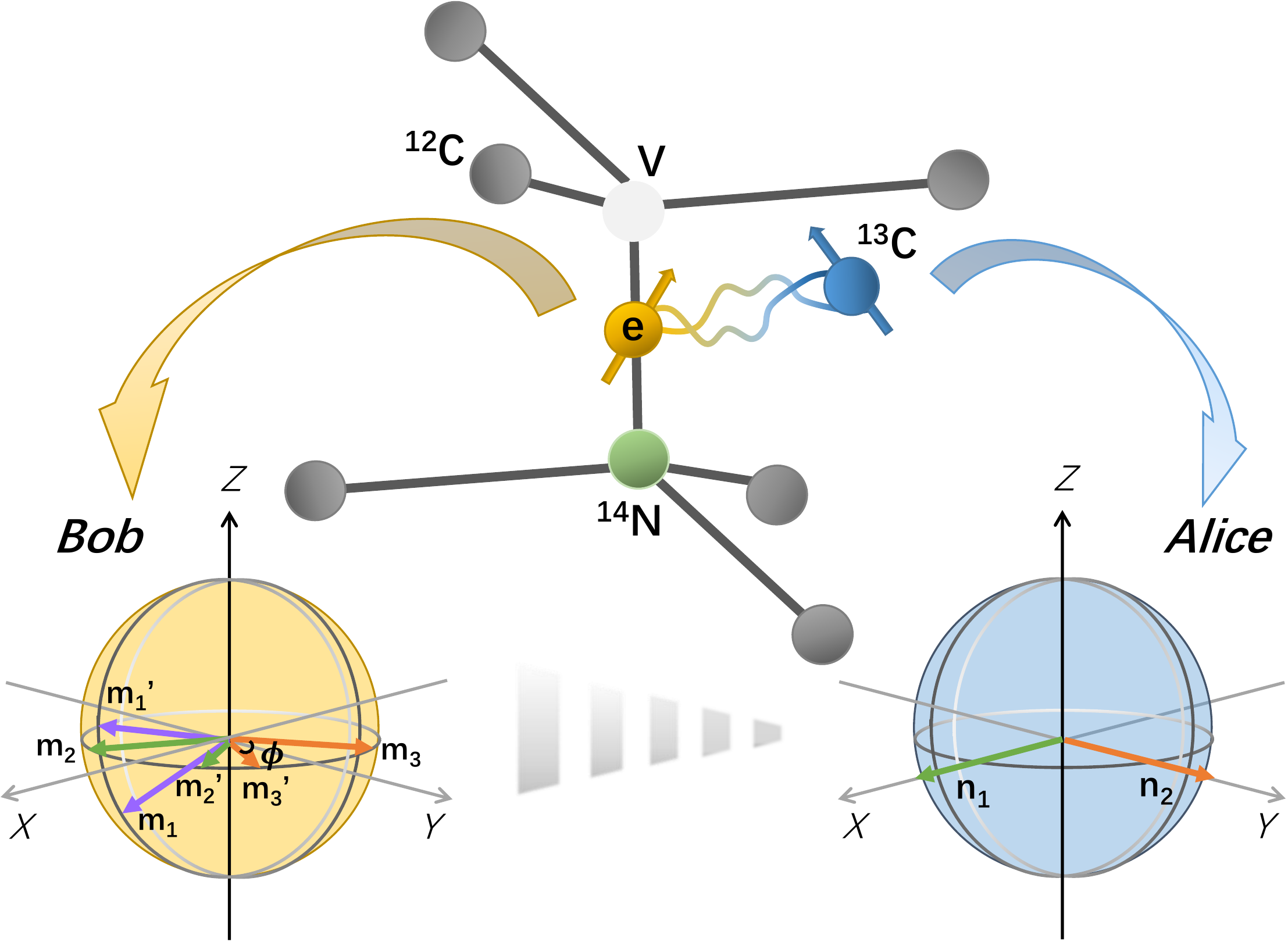}
	\caption{Tests of Leggett's nonlocal hidden-variable theories with solid-state spins in a diamond nitrogen-vacancy center. An electron spin of the nitrogen-vacancy center is prepared to be entangled with one of its surrounding $^{13}C$ nuclear spins. We denote the nuclear spin and the electron spin by Alice and Bob, respectively. Measurement settings on the Poincar$\acute{e}$ sphere are shown for testing the Ineq. (\ref{Ineq26}).}
	\label{Fig:FigAliceBob}
\end{figure}

\begin{figure*}[!t]
	\includegraphics[trim=0cm 0cm 0cm 0cm, clip,scale=0.55]{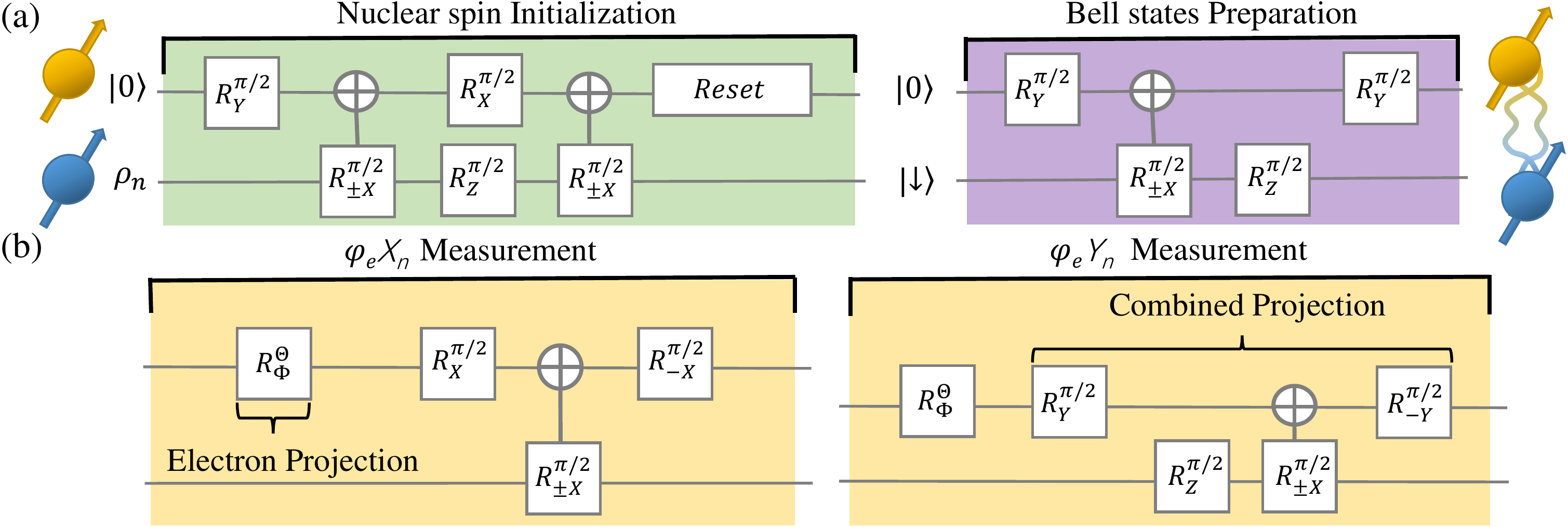}
	\caption{(a) Experimental sequences for initialization of a weakly coupled $^{13}C$ nuclear spin state and preparation of one of the electron-nuclear Bell states. The initial state of nuclear spin can be flipped by tunning the quantum gate parameters and the phase settings of two electron microwave $\pi$ pulses in entanglement preparation can be changed to transform generation between four Bell states. (b) The measurement sequences for jointly measuring the electron and nuclear spins, in a rotated  electron-nuclear basis $\varphi_eX_n$/$\varphi_eY_n$.  After appropriate basis rotation, the corresponding $\varphi$ settings electron spin state is projected to $X_e$/$Y_e$ axis and the joint detection can be achieved by following combined projective optical readout \cite{supp-NVLI}.}
	\label{Fig:Sequence}
\end{figure*}

In this paper, we fill this important gap by reporting an experimental test of Leggett's NLHV model with solid-state spins in a diamond nitrogen-vacancy center (see Fig.\ref{Fig:FigAliceBob} for a pictorial illustration). Following Branciard {\it et al.} \cite{branciard2008testing}, we derive two Leggett-type of inequalities with six and eight measurement settings respectively,  without assuming a time ordering of the events as in Leggett's original paper \cite{leggett2003nonlocal}.  We entangle an electron spin in the NV center with a surrounding $^{13}C$ nuclear spin to form a maximally entangled Bell state with high fidelity and perform appropriate single-shot projection measurements on the electron spin to measure the correlations between the electron and nuclear spins. We observe that for large measurement parameter regions, the entangled states violate both the derived six- and eight-setting Leggett-type inequalities.  In particular, for the six-setting (eight-setting) inequality, the maximal violation  exceeds the classical bound by $4.0$ and $10.9$ ($7.1$ and $15.5$) standard deviations for the raw data and the data after correction of the readout error, respectively.  Our experiment results are in full agreement with quantum predictions and thus falsify Leggett's NLHV model in a solid-state system.

To begin with, let us first briefly introduce two inequalities based on Leggett's nonlocal model, following a simpler approach introduced  by Branciard {\it et al.} \cite{branciard2008testing}. Consider a common Bell-type experimental scenario: two observers, denoted by Alice and Bob, perform measurements labeled by \textbf{n} and \textbf{m} on their qubits, respectively. The outcomes are denoted by $\alpha$ and $\beta$ ($\alpha,\beta=\pm1$). For the qubit case, \textbf{n} and \textbf{m} are unit vectors on the Poincar$\acute{e}$ sphere (see Fig.\ref{Fig:FigAliceBob} for an illustration) and are independently and freely chosen by Alice and Bob. According to hidden-variable theories \cite{Clauser1969Proposed}, the conditional probability distribution $P(\alpha,\beta|\bm{n,m})$ can be decomposed into a statistical mixture of correlations characterized by the hidden variable $\lambda$:
\begin{equation}
P(\alpha,\beta|\bm{n,m}) = \int_{\Gamma}\rho(\lambda)P_{\lambda}(\alpha,\beta|\bm{n,m})d\lambda
\end{equation}
where $\Gamma$ is the total $\lambda$ space and $\rho(\lambda)$ is a statistical distribution of $\lambda$ which satisfies $\rho(\lambda)\ge0$ and$\int_{\Gamma}\rho(\lambda)d\lambda=1$. Strikingly, the constraint of non-negativity of probabilities $P_{\lambda}(\alpha,\beta|\bm{n,m})\geq 0$ is sufficient to derive testable Leggett-type inequalities which are satisfied by Leggett's NLHV model but can be violated by quantum predictions. The simplest inequality reads (see \cite{branciard2008testing} and \cite{supp-NVLI} for details):
\begin{eqnarray}
\mathcal{I}_{26}(\phi) &\equiv & |C_{\mathbf{n}_1,\mathbf{m}_1}+C_{\mathbf{n}_1,\mathbf{m}'_1}|+|C_{\mathbf{n}_1,\mathbf{m}_2}+C_{\mathbf{n}_1,\mathbf{m}'_2}| \nonumber \\
&+ &|C_{\mathbf{n}_2,\mathbf{m}_3}+C_{\mathbf{n}_2,\mathbf{m}'_3}| + 2\sin \frac{\phi}{2}\leq 6, \label{Ineq26}
\end{eqnarray}
where $C_{\mathbf{n},\mathbf{m}}=\sum_{\alpha,\beta}\alpha\beta P(\alpha,\beta|\bm{n,m})$ denotes the usual correlation function and $\phi$ is the angle between a pair of vectors $\mathbf{m}_i$ and $\mathbf{m}'_i$ $(i=1,2,3)$ \cite{supp-NVLI}. We note that the above inequality is a bit different from the original one introduced in Ref. \cite{branciard2008testing}, where three measurement settings for Alice's side were used. Here, we use only two settings for Alice because the tests of this modified inequality is easier to implement in our NV experimental setup.

In the experiment, we use solid-state spins, namely an electron spin and a surrounding $^{13}C$ nuclear spins in a diamond NV center \cite{doherty2013nitrogen}, to test Leggett-type inequalities.  The NV center is a natural doped structure which is composed of a vacancy and an adjacent nitrogen atom that replace the two neighboring carbon atoms \cite{neumann2012towards}. Our experiments utilize the negative charge state of the NV center with an electron spin $S=1$ (denoted as $|m_s=\pm1\rangle$ and $|m_s=0\rangle$) and a nearby weakly coupled $^{13}C$ nuclear spin $I=1/2$ (denoted as $|\uparrow\rangle$ and $|\downarrow\rangle$) in a cryostat at temperature around $8$ K, with optical initialization and readout achieved through use of resonant transitions \cite{bernien2014control} between excited states and ground states.

To realize efficient multi-qubit control, we need to design a set of single-qubit gates and electron-nuclear two-qubit entangling gates. With a magnetic field $B_z$ aligned along the NV symmetry axis and under the rotating wave approximation, the effective Hamiltonian of the system in the rotating frame with respect to the modulated electron energy splitting describing the electron spin and a single $^{13}C$ nuclear spin has the form
\begin{equation}
H_{eff} = A_{zz}\hat{S}_z\hat{I}_z+A_{zx}\hat{S}_z\hat{I}_x+\gamma_nB_z\hat{I}_z,
\end{equation}
Where $\hat{S}_z=\text{diag}\{1,0,-1\}$ denotes the $z$-component of the spin-one operator, and $\hat{I}_x$ and $\hat{I}_z$ are the Pauli-X and Pauli-Z matrix, respectively (here we define NV symmetry axis as the $z$ axis); $\gamma_n$ is the gyromagnetic ratio of the $^{13}C$ nuclear spin; $A_{zz}$ and $A_{zx}$ form the parallel and perpendicular components of the hyperfine interaction term between the electron spin and the nuclear spin with their values determined precisely from previous experiments \cite{hou2019experimental}. Due to the particular mutual interaction, the $^{13}C$ nuclear spin processes around the axis conditioned on the electron spin state, so we can construct a set of selective electron-nuclear two-qubit gates based on the dynamical decoupling sequences \cite{taminiau2012detection}.

\begin{figure}[!t]
	\includegraphics[trim=0cm 0cm 0cm 0cm, clip,width=0.98\linewidth]{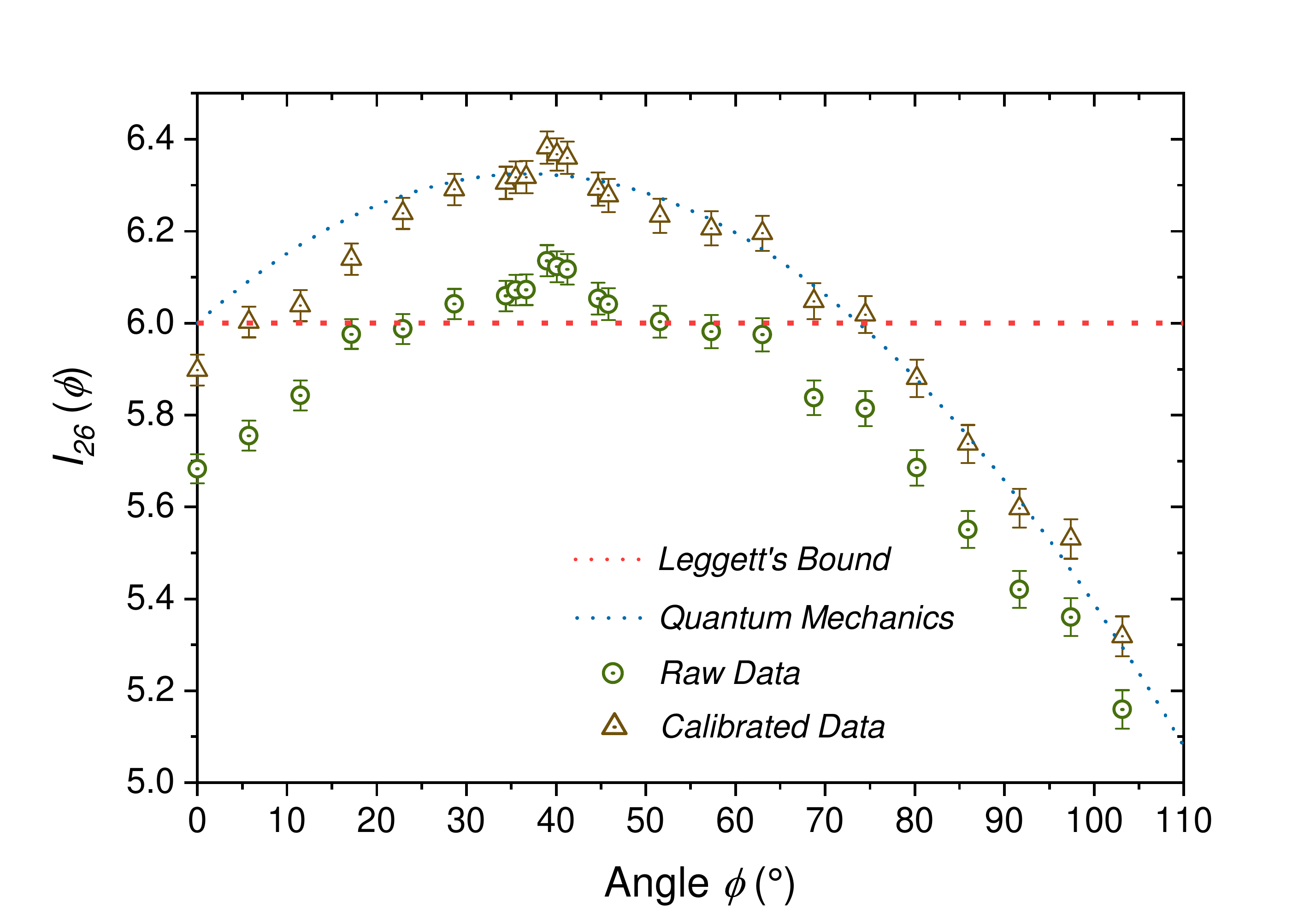}
	\caption{Experimental violations of the Leggett's inequality  with six measurement settings. 	
The dashed orange line  indicates the bound of inequality (\ref{Ineq26}), which is satisfied by Leggett's non-local hidden variable model.  The dotted line denotes the quantum-mechanical prediction and the region above the dashed line implies quantum violation.  Our experimental raw data (denoted by circles) of $\mathcal{I}_{26}$ exceed the Leggett's bound for $23.02^{\circ}<\phi<51.57^{\circ}$. The largest violation is observed for $\phi_\emph{{max}}=38.96^{\circ}$. The triangles show the data after correction of the readout error, and the violating $\phi$ region in the case becomes broader and the maximal violation also gets bigger. Here, the errorbars denote the readout standard deviations.}
	\label{Result:I26}
\end{figure}

One can verify that the inequality (\ref{Ineq26}) can be violated in quantum mechanics for a range of $\phi$ and for various quantum entangled states. The maximal violation is achieved when the left side of Eq. (2) equals
$\sqrt{40}$ and this happens at $\phi_0 =2\arctan\frac{1}{3}
\approx36.9^\circ$  under the maximally entangled singlet state \cite{branciard2007experimental}. For this optimally chosen setting, the minimal visibility to observed violation of the inequality (2) is about $V_{min}=94.3\%$. The minimal visibility measures how much white noise can be added into the Bell state so that the inequality is still violated, i.e., the minimal value of $V$ under the condition that the state $\rho_V=V|\Phi^-\rangle\langle\Phi^-|+(1-
V)\bm{I}/4$ (here $\bm{I}$ is the $4\times4$ identity matrix) violates the inequality (\ref{Result:I26}). The corresponding minimal fidelity is estimated to be $F_{min}=\sqrt{3V_{min}+1}/2\approx97.8\%$. This high value of the required minimum fidelity is a significant challenge for an experimental observation of violation of the Leggett inequality. In our experiments, we have used the dynamical decoupling \cite{taminiau2012detection} and optimized the sequence parameters to meet this requirement \cite{supp-NVLI}.

We choose the electron spin to act as the control qubit and the $^{13}C$ nuclear spin undergoing the conditional rotation as the target qubit. To protect single nuclear spin from the decoherence effect and avoid the unwanted crosstalk between multiple nuclear spins, we need to optimize the parameters of single and controlled quantum gates based on the precisely characterized hyperfine interaction couplings \cite{hou2019experimental}. The sequence to achieve the nuclear spin state initialization and electron-nuclear spin entanglement \cite{reiserer2016robust} is shown in Fig.\ref{Fig:Sequence}(a). First we prepare the electron spin in $|0\rangle$ and after a swapping procedure \cite{taminiau2014universal}, the nuclear spin is initialized onto $|\uparrow\rangle$ or $|\downarrow\rangle$ determined by the controlled quantum gate parameters. We then reset the electron spin to be on state $|0\rangle$ and apply entangling gate on the electron and nuclear spins. After this, the electron and nuclear spins are maximally entangled with the state $|\Phi^-\rangle=(|0_e\uparrow_n\rangle -|1_e\downarrow_n\rangle)/\sqrt{2}$. From the measured expectation values of entanglement witness operators described in Ref. \cite{toth2005entanglement}, we can obtain a lower bound on the fidelity of our prepared Bell state at $98.2(5)\%$.

For the quantum state measurements on the entangled Bell state, instead of measuring electron spin and nuclear spin separately \cite{cramer2016repeated}, we choose appropriate basis projections of both electron and nuclear spin followed by Z-basis optical measurement to readout electron-nuclear spin state simultaneously in a single-shot readout scheme with sufficient average fidelity at cryogenic temperature [Fig.\ref{Fig:Sequence}(b)]. The specific process starts with the rotation of electron-spin from pairs ($\bm{m_i,m_i'}$) which satisfy corresponding $\varphi$ settings to the $X_e/Y_e$ basis of the two-qubit system followed by a joint measurement of $X_eX_n/Y_eY_n$ using the sequence particularly designed (see  \cite{supp-NVLI} for the details). In this way, we are able to measure the desired correlations appearing in Ineq. (\ref{Ineq26}).  

We vary the measurement angle parameter $\phi$ in a discrete way and for each value of $\phi$ we  measure the quantum expectation value of $\mathcal{I}_{26}$.  Our experimental results are shown in Fig.\ref{Result:I26}. From this figure, our experimental results match the theoretical predictions qualitatively and the Leggett's inequality (\ref{Ineq26}) is violated for  $23.02^{\circ}<\phi<51.57^{\circ}$. The largest violation occurs at $\phi_\emph{{max}}=38.96^{\circ}$ and the violation for experimental raw data is $6.136\pm 0.034$, violating the Ineq. (\ref{Ineq26}) by more than four standard deviations.  In addition, since in our experiments the state initialization and projective  readout procedures consist of the same number of similar gate operations, we can follow a standard recipe as in Ref.  \cite{cramer2016repeated} to correct the readout error (see also \cite{supp-NVLI} for the details). After this correction, each measured correlation will be more close to its corresponding theoretical prediction without experimental imperfections and the violation of the Leggett's inequality will be further enhanced. Our experimental data after the readout error correction is also shown in
in Fig.\ref{Result:I26}. It is clear from this figure that after the readout error correction our experimental results agree excellently with the theoretical quantum predictions and the  the maximal violation in this case is $6.382\pm0.035$, which violate the Ineq. (\ref{Ineq26}) by more than $10.9$ standard deviations.

\begin{figure}[!t]
\centering
	\includegraphics[trim=0cm 0cm 0cm 0cm, clip,width=0.98\linewidth]{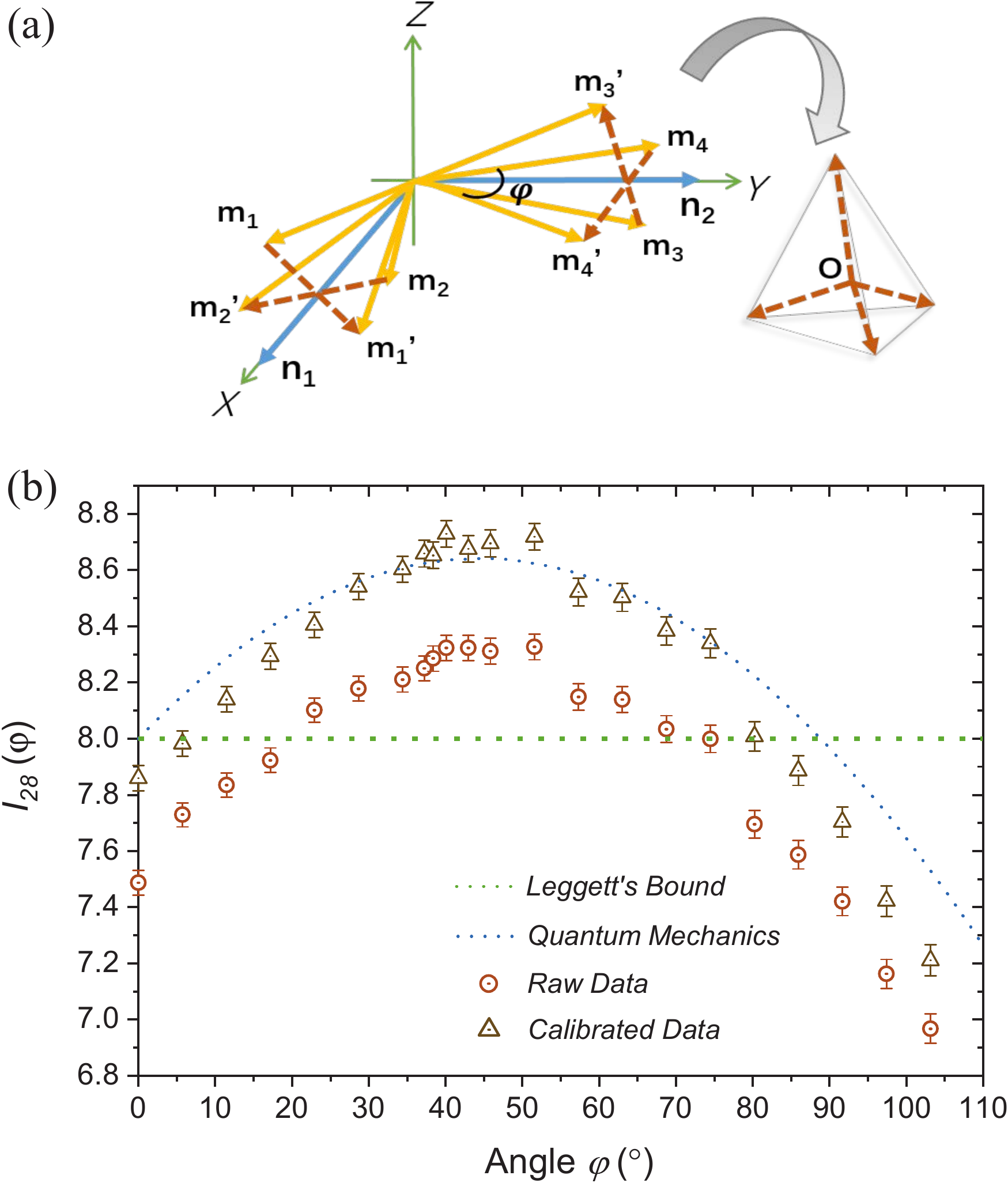}
	\caption{(a) Measurement settings for nuclear (blue arrows) and electron (yellow arrows) spins on the Poincar$\acute{e}$ sphere for testing the Ineq. (\ref{Ineq28}) \cite{supp-NVLI}. The four vectors $\{\vec{e_i}=\bm{m_i-m_i'}\}$ point to the four vertices of the regular tetrahedron. 	(b) Experimental violations of the Leggett's inequality (\ref{Ineq28}). The quantum expectation of $\mathcal{I}_{28}$ exceeds the Leggett's bound for $17.96^{\circ}<\varphi<71.34^{\circ}$ and the maximal violation occurs at $\varphi_\emph{{max}}=40.11^{\circ}$ for raw data (circle). With readout correction , we get the broader violating $\phi$ region and the bigger maximal violation (triangle).}
	\label{Result:I28}
\end{figure}

To obtain stronger quantum violations, one can increase the number of measurement settings, similar to the case of testing Bell inequalities. In Ineq. (\ref{Ineq26}),  Alice has two measurement settings and Bob has six settings. Following a similar derivation, we obtain another Leggett-type inequality, where Bob has eight measurement settings  \cite{supp-NVLI}:
\begin{eqnarray}
\mathcal{I}_{28}(\varphi) &\equiv & |C_{\mathbf{n}_1,\mathbf{m}_1}+C_{\mathbf{n}_1,\mathbf{m}'_1}|+|C_{\mathbf{n}_1,\mathbf{m}_2}+C_{\mathbf{n}_1,\mathbf{m}'_2}| \nonumber \\
& +&|C_{\mathbf{n}_2,\mathbf{m}_3}+C_{\mathbf{n}_2,\mathbf{m}'_3}| + |C_{\mathbf{n}_2,\mathbf{m}_4}+C_{\mathbf{n}_2,\mathbf{m}'_4}| \nonumber \\
&+& \frac{8}{\sqrt{6}}\sin\frac{\varphi}{2}\le 8. \label{Ineq28}
\end{eqnarray}
It is straightforward to check that the inequality (\ref{Ineq28}) is violated in quantum physics for a large range of $\varphi$ with measurement settings shown in Fig.\ref{Result:I28}(a),
and the maximal violation occurs at $\varphi_0 =\pi-2\arctan\sqrt{\frac{6}{7}}
\approx44.4^\circ$. The maximal violation of inequality (\ref{Ineq28}) is $8\sqrt{\frac{7}{6}}$ and the threshold visibility is $V_{min}\approx91.3\%$ with the corresponding  threshold fidelity $F_{min}=96.7\%$, which are both smaller than the ones for the six-setting Ineq. (\ref{Ineq26}). 
Our experimental results for testing Ineq.(\ref{Ineq28}) is plotted in Fig.\ref{Result:I28}(b), from which it is evident that the inequality is violated for $17.96^{\circ}<\varphi<71.34^{\circ}$. The maximal violation is $8.323 \pm 0.045$,  occurring at $\varphi_\emph{{max}}=40.11^{\circ}$, which violates the Ineq. (\ref{Ineq28}) by more than $7.1$ standard deviations. With readout error correction, the maximal violation is $ 8.729 \pm 0.047$ \cite{supp-NVLI}, violating the Ineq. (\ref{Ineq28}) by more than $15.5$ standard deviations. Our experiments confirm that the violation of the eight-setting inequality (\ref{Ineq28}) is notably larger than that of the six-setting inequality (\ref{Ineq26}).

In summary, we have experimentally tested Leggett's NLHV model in a solid-state system for the first time. Our experimental results are in agreement with quantum predictions and show clear violation of the Leggett-type inequalities with a high level confidence, thus falsifying Leggett's model with solid-state spins. Our discussion is mainly focused on the two-qubit case, but its generalizations to multiple qubits are possible and worth future investigations. In addition, it would also be interesting to experimentally test nonlocal causality \cite{ringbauer2016experimental} with solid-state spins in a similar setup.


%
%
%
%
%
%


\begin{acknowledgments}
This work was supported by the Frontier Science Center for Quantum Information of the Ministry of Education of China, Tsinghua University Initiative Scientific Research Program, and the National key Research and Development Program of China (2016YFA0301902).
\end{acknowledgments}

\bibliographystyle{apsrev4-1-title}
\bibliography{Dengbib,Leggett_Ineq}

\end{document}